\documentclass{article}
\usepackage{spconf,amsmath,graphicx,array,subfig}

\newcolumntype{T}[1]{%
    >{\centering\arraybackslash\hspace{0pt}}p{#1}}%

\title{An ANALYSIS OF DEGENERATING SPEECH DUE TO PROGRESSIVE DYSARTHRIA ON ASR PERFORMANCE}

\name{Katrin Tomanek$^1$, Katie Seaver$^1$, Pan-Pan Jiang$^1$, Richard Cave$^1$$^,$$^3$, Lauren Harrell$^1$, Jordan R. Green$^1$$^,$$^2$}
\address{$^1$Google LLC, $^2$MGH Institute of Health Professions, $^3$Language and Cognition, UCL  \\
 \texttt{katrintomanek@google.com, jgreen2@mghihp.edu}}

\begin{document}
%
\maketitle
\begin{abstract}
Although personalized automatic speech recognition (ASR) models have recently been designed to recognize even severely impaired speech, model performance may degrade over time for persons with degenerating speech. The aims of this study were to (1) analyze the change of performance of ASR over time in individuals with degrading speech, and (2) explore mitigation strategies to optimize recognition throughout disease progression. Speech was recorded by four individuals with degrading speech due to amyotrophic lateral sclerosis (ALS). Word error rates (WER) across recording sessions were computed for three ASR models: Unadapted Speaker Independent (U-SI), Adapted Speaker Independent (A-SI), and Adapted Speaker Dependent (A-SD or personalized). The performance of all three models degraded significantly over time as speech became more impaired, but the performance of the A-SD model improved markedly when it was updated with recordings from the severe stages of speech progression. Recording additional utterances early in the disease before speech degraded significantly did not improve the performance of A-SD models. Overall, our findings emphasize the importance of continuous recording (and model retraining) when providing personalized models for individuals with progressive speech impairments. 
\end{abstract}

\section{Introduction}
\label{sec:intro}

Amyotrophic lateral sclerosis (ALS), also known as motor neuron disease is a progressive, ultimately fatal disease causing progressive loss of motor function~\cite{Oliver2017}. ALS progression is heterogeneous in terms of the pattern of spread across body parts and the rate of functional decline \cite{green2021}. Between 80-95\% of people living with ALS (PALS) experience progressive dysarthria and increasing difficulty communicating daily needs via speech~\cite{Beukelman2011}. Speech decline is fastest for individuals first presenting symptoms in the head and neck muscles~\cite{Eshghi2022,Makkonen2018} and dysarthria can progress rapidly, rendering speech unusable within 23 months from diagnosis~\cite{Eshghi2022}.

Automatic speech recognition (ASR) may significantly extend functional communication in PALS ~\cite{Cave2021}. However, the speech of PALS may be challenging to recognize due to progressing dysarthria~\cite{Caballero2014}. Dysarthria due to ALS is characterized by spectral and temporal alterations to the speech signal resulting in prolonged, distorted, and less distinct phonemes~\cite{Rowe2022}, increased nasal resonances~\cite{Eshghi2021}, decreased vocal harmonics~\cite{tomik2015}, and increased duration and frequency of pauses~\cite{green2004}.

Recent work shows that ASR systems trained on typical speech poorly generalize to dysarthric speech~\cite{DeRussis2019}. In contrast, personalized models trained using samples from the end-user speaker, can be highly accurate - even for severe dysarthria~\cite{green2021, Shor2019, doshi2021} under some speaking conditions (i.e., short, prompted phrases) and with limited amount of data to personalize on~\cite{tobin2022}. However the performance of these models is likely to degrade over time in PALS as speech becomes slower and less intelligible. Little is known about the tolerance of personalized ASR models to progressive speech changes, and when models need to be updated to optimize accuracy. Specialized training strategies and recording schedules may be needed to boost performance during advanced disease progression. Performance might be enhanced by using recordings collected during the early stage of progression for training.  For this study, we  identified four speakers from the Euphonia Corpus~\cite{macdonald2021} where patterns of degenerating speech could be observed. We then analyzed how speaker independent and speaker dependent ASR models degrade over time as a function of speech severity, and explored strategies to improve personalized models over the course of progression with limited amounts of new data.

\section{Methods}
\label{sec:methods}

\begin{table*}[t]
  \centering
  \small
  \begin{tabular}{|l|l|l|l|l|l|}
  \hline
  & & bin 1 & bin 2 & bin 3 & bin 4\\
  \hline 
  Subject 1 & \# test utterances & 1211 & 45 & 213 & 487 \\
  & severity & MILD & MODERATE & SEVERE &  PROFOUND \\
  & days from baseline & 0 - 55 & 83 - 89 & 191 - 216 & 324 - 421 \\
  \hline
  Subject 2 & \# test utterances & 61 & 100 & 295 & 49 \\
  & severity & MODERATE & SEVERE & PROFOUND & ANARTHRIC\\
  & days from baseline & 0 - 14 & 17 - 56 & 42 - 132 & 133 - 194 \\
  \hline
  Subject 3 & \# test utterances & 262 & 121 & 48 &  \\
  & severity & MILD & MODERATE & SEVERE  & \\
  & days from baseline & 0 - 29 & 191 - 292 & 314 - 314 &  \\
  \hline
  Subject 4 & \# test utterances & 387 & 233 & 88 & \\
  & severity & MILD & MODERATE & SEVERE & \\
  & days from baseline & 0 - 50 & 55 - 79 & 214 - 220 & \\
  
  \hline
  \end{tabular}
  \caption{Severity bins per speaker and resulting test set sizes.}
  \label{tab:severity_bins}
\end{table*}
\vspace{-1em}

\subsection{Subjects and speech recordings}

Four subjects with progressive dysarthria were identified from the Euphonia dataset, a corpus of over 1 million speech samples from over 1000 individuals with impaired speech ~\cite{macdonald2021}. The Euphonia dataset was collected over several years and many of the subjects recorded over multiple months, allowing us to find cases with declining speech. The four subjects who were selected had (1) at least a 10\% drop in ASR performance on the U-SI model over time (Section~\ref{ss:asr_models}), and (2) an increase in speech severity by at least two points between their first and last recording sessions (Section~\ref{ss:severity_rating}). Speech recordings
were binned into successive 30 day intervals so that, for example, speech recorded between the first and 30th day
were coded as bin 1, and data recorded between the 31st and 60th day from first recording were coded as bin 2. After the speech severity ratings were first assigned to each 30-day bin, we then re-grouped these into fewer, purely severity based bins. For example, for a speaker with a mild severity rating across two consecutive 30-day bins, all recordings were labeled as “mild” and collapsed into one bin. We obtained 3-4 severity-based bins per speaker (Table~\ref{tab:severity_bins}).

\subsection{Perceptual speech severity ratings}
\label{ss:severity_rating}

Speech severity rating was completed by two licensed speech-language pathologists (SLP), who listened to at least 10 utterances from each original 30-day bin and rated overall speech severity on a 5-point Likert scale (typical, mild, moderate, severe, and profound) ~\cite{stipancic2021}. The raters used professional grade headphones and were allowed to adjust the gain as needed. Interrater reliability was assessed by computing a two-way, single measure model intraclass correlation coefficient
which resulted in an ICC of 0.88~\cite{koo2016}. For the reliability analysis, the two SLPs rated speech severity for the same 50 recordings on a different dataset, part of the parent project.

\subsection{Selecting Utterances for experimentation}

The Euphonia dataset consists of recordings from different domains\footnote{e.g., home automation, caregiver requests, and conversational phrases} which vary in average utterance length. To control confounding effects of utterance length we include only short phrases of 3-5 words in length. Such short phrases were chosen because those are most prevalent in the Euphonia dataset and we wanted to maximize the number of utterances per speaker (see Table~\ref{tab:severity_bins} for a summary). Because the goal of this analysis was to compare model performance across different levels of speech degradation, we held the number of training data per speaker constant in order to not confound the model comparisons by the total volume of training data (see ~\cite{tobin2022} on how training set size influences model personalization on impaired speech). We chose the training set sizes of each speaker to be as large as possible while leaving enough utterances per bin to be used as test set. These training set sizes can be thought of as a “budget” (i.e., the maximum number of utterances we assume a speaker to record). Table~\ref{tab:training_data} shows the training set size per speaker. 

Training sets of the identified size were then randomly sampled from the recordings from each severity bin. From the remaining utterances we created test sets for each speaker, ensuring no phrase overlap with their training sets. The same resulting test sets per speaker and per severity bin were used across all experiments ensuring comparability (see Table~\ref{tab:severity_bins}).

\begin{table}[h]
  \centering
    \small
  \begin{tabular}{|l|l|l|l|}
  \hline
  Subject 1 & Subject 2 & Subject 3 & Subject 4\\
  \hline
  400 & 100 & 300 & 300 \\
  \hline
  \end{tabular}
  \caption{Number of training utterances per subject.}
  \label{tab:training_data}
\end{table}
\vspace{-1.4em}

\subsection{Evaluation}

We calculated word error rate (WER) per severity bin and applied bootstrap sampling (1000 repetitions with replacement) to obtain estimates of the mean WER as well as 95\% confidence intervals approximated by +/- 2 standard deviations.\footnote{Analysis showed that the WER across samples was normally distributed.}
For comparing personalized models, 95\% confidence intervals (CIs) for the difference in WER within speaker were generated using bootstrap sampling as well, where CIs not overlapping 0 show significant difference between two strategies.

\subsection{ASR Models}
\label{ss:asr_models}

We used end-to-end ASR models based on the well-studied RNN-T architecture~\cite{graves2013sequence}, with an encoder network consisting of 8 layers and the predictor network  of 2 layers of uni-directional LSTM cells. Inputs were 80-dimensional log-mel filterbank energies. Outputs were probability distributions over a 4k word piece model vocabulary. 

The Unadapted Speaker Independent (U-SI) model was trained as described in~\cite{narayanan2018}, using $\approx 162k$ hours of typical speech (from Google's internal production dataset).
For the Adapted  Speaker  Independent (A-SI) model, we further fine-tuned the U-SI model on a large subset of the whole Euphonia dataset with the goal to provide a model that should work better out-of-the-box for impaired speech. For this, the Euphonia dataset was split into a training and a test set across all speakers. There was no speaker- or phrase-overlap between these two sets and also excluded the speakers used in this study from the training portion. The training portion consisted of $\approx 1200$ speakers of a wide range of speech impairment types and severities (including $\approx 20\%$ ALS), and included $\approx 900k$ utterances from various domains.
Lastly, the Adapted Speaker Dependent (A-SD) models were also adaptations of the U-SI model, but fine-tuned only with data from the specific speaker. We followed the general personalization recipe as described in \cite{green2021}. We explored variants of the A-SD model based on sampling data from different severity bins.

\section{Results}
\label{sec:results}

\subsection{Impact of Degenerating Speech on ASR performance}
\label{ss:impact}

\begin{figure*}[t]
    \centering
    \subfloat{\includegraphics[width=2.9in]{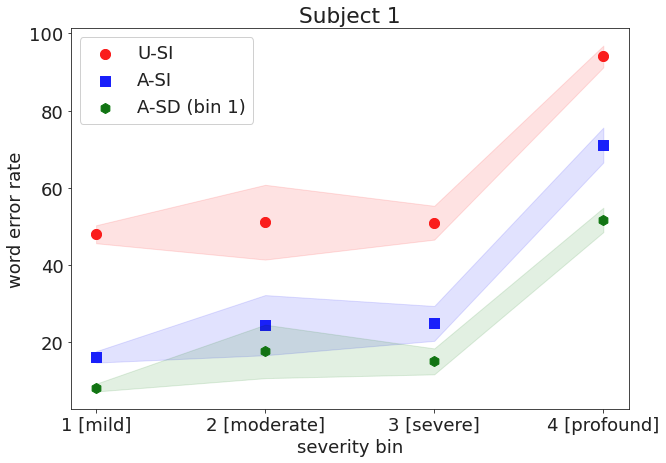}}
    \subfloat{\includegraphics[width=2.9in]{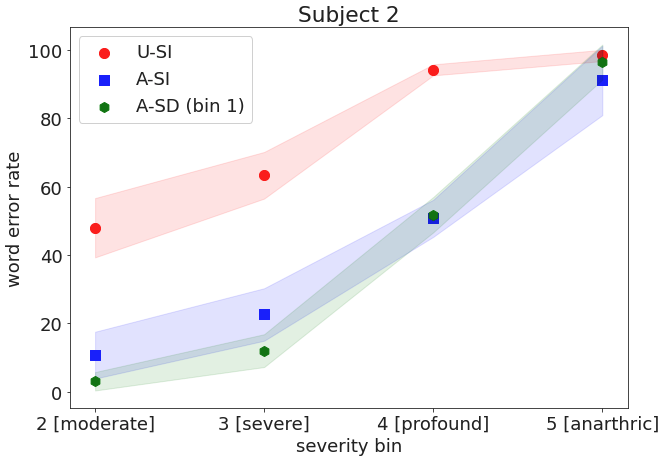}}\\
    \subfloat{\includegraphics[width=2.9in]{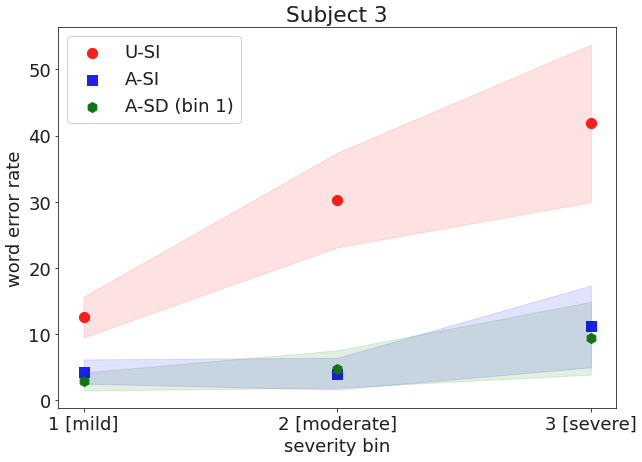}}
    \subfloat{\includegraphics[width=2.9in]{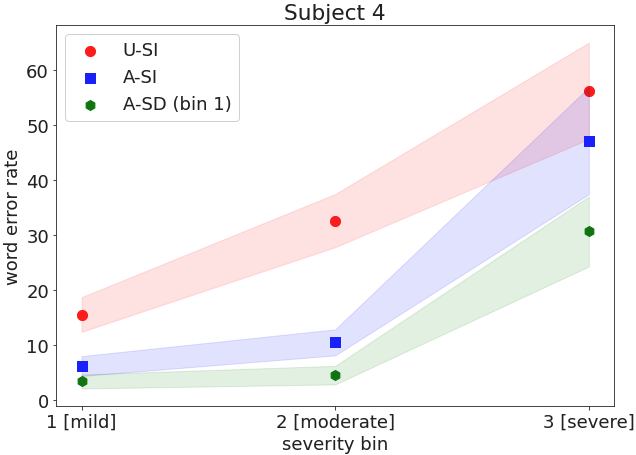}}
    \caption{WER across all severity bins and 3 different models for all four speakers. Each dot is the mean WER on the speaker’s test set with the given model and the shaded area represents the 95\% CIs.}
    \label{fig:wer_across_bins}
\end{figure*}

We first compared the performance of the U-SI, the A-SI and the A-SD model personalized on the training data of the first severity bin per speaker to analyze the impact of degenerating speech on recognition performance. Figure~\ref{fig:wer_across_bins} displays a progression charts. 
As expected, the U-SI models consistently performed the worst across all speakers and severity levels while the A-SD models performed best and the A-SI models often were somewhere in between. As a general observation, an increase in severity consistently led to an increase in WER, especially in the higher severity levels.

For Subject 2, recognition performance of the A-SD model was almost as poor as the U-SI model; in this case, even the A-SI performed slightly better. For Subject 3, the A-SI was not much worse than the A-SD model. Note that the A-SI model did not include any data from the target speaker.

\subsection{Mitigation Strategies}

\begin{figure}[t]
    \centering
    \includegraphics[width=0.45\textwidth]{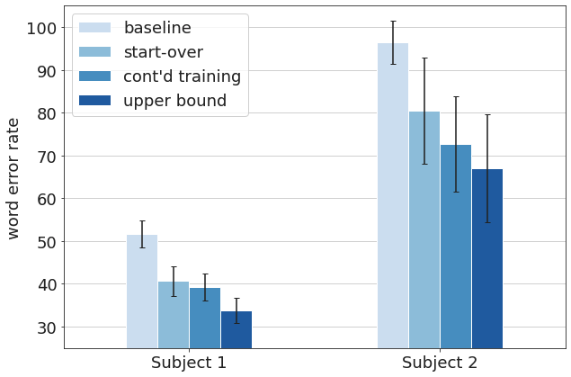}
    \caption{WERs on the highest severity bin for A-SD models.}
    \label{fig:mitigatio_strategies}
\end{figure}

Based on our findings that the performance of personalized models degraded when trained on speech recorded during the early stage of speech impairment,  we explored the effectiveness of two practical mitigation strategies for optimizing recognition during the most severe stages of speech decline when fewer training samples are typically available. In these experiments, the overall maximum number of training utterances was held constant (Table~\ref{tab:training_data}), and we only tested the performance of the A-SD models on the last severity bin, when WERs are highest. We compared the following 4 variants of the A-SD model:

\begin{itemize}
    \item Baseline - 100\% of data from bin 1 (as in Section~\ref{ss:impact})
    \item Mitigation Strategy 1 (“start-over”) - Assuming we have used 50\% of our recording budget on bin 1, we now use another 50\% of recordings from the last severity bin. In the start-over scenario, we use only this 2nd 50\% of recordings for adaptation.
    \item Mitigation Strategy 2 (“Continued Training”) - data allocation like in the start-over scenario but here we simulate training continuation in that we use both the bin 1 and bin 4 recordings for adaptation.
    \item Upper Bound - 100\% of the recordings from bin 4 – this is an idealized and unpractical scenario only used to show the best possible performance if all training data is from the most recent severity level.
\end{itemize}
Figure~\ref{fig:mitigatio_strategies} shows results for Subject 1 and Subject 2.\footnote{Other subjects omitted due to insufficient utterances and severity stages to allow for meaningful experiments.} The baseline model performed the worst of the four scenarios, while  training on as much data as possible from the last severity phase resulted in the best recognition. While these findings were expected, they clearly show the negative impact of “outdated” data and how much, in contrast, a model can be improved by using the most recent data of the same size.\footnote{Note that for both speakers, the final WERs were high even when data from the most recent severity bin was included. This can be attributed to the relatively small number of recordings in these bins (Table~\ref{tab:severity_bins}). In a typical recording scenario, one would probably want to increase the recording budgets for speakers with such severe levels of speech impairment.}

Comparing the two mitigation strategies (start-over and continued training), we found both to significantly improve WER over the baseline approach for both subjects. The continued training strategy provided a significant improvement over the start-over strategy for Subject 2 but not for Subject 1. The upper bound scenario was significantly better than both mitigation strategies for both subjects, which emphasizes that doubling the number of recordings in later stages may be beneficial (if larger amounts of recordings are possible).  Table~\ref{tab:mitigation_ci} shows the 95\% CIs for these comparisons.

\begin{table}[h]
{  \centering
    \footnotesize
\begin{tabular}{c | T{0.04\textwidth} | T{0.04\textwidth} |T{0.04\textwidth} | T{0.045\textwidth} |T{0.045\textwidth} |T{0.045\textwidth} }
 Subject    & Baseline - Upper Bound & Baseline - Start Over & Baseline - Cont'd & Upper Bound - Start Over & Upper Bound - Cont'd & Cont'd - Start Over \\ \hline
1 & 19.76, 39.42 & 6.92, 25.03 & 14.37, 31.75 & -22.60, -4.63 & -12.36, -0.69 & -12.82, -1.35 \\ \hline
 2 & 15.71, 19.93 & 8.53, 12.87 & 10.35, 14.67 & -8.94, -5.30 & -7.02, -3.60 & -3.65, 0.03
\end{tabular}
  \caption{95\% Confidence Intervals for the difference in WER between mitigation strategies}
  \label{tab:mitigation_ci}
}
\end{table}

\section{Discussion}
\label{sec:discussion}

This study investigated the impact of degrading speech on ASR accuracy in individuals with progressive dysarthria. Speech samples were recorded over time and selected to represent a substantial within-subject decline in speech. Recognition accuracy of the three ASR models decreased as speech degraded, particularly during the more severe stages of speech decline. 
To the best of our knowledge, this is the first time that the impact of speech degeneration on personalized models has been studied systematically.

Our experiments suggest that personalized models become less effective over the course of progression unless updated with more current recordings. Both the start-over scenario discarding “outdated” recordings, as well as continued training adding more recent and keeping outdated data, significantly improved recognition when compared to recording all utterances up-front without updating. Our experiments also suggest, that in absence of more recent data, keeping data from previous severity stages
does not seem to incur any harm, but potentially improves performance.

Overall, our findings emphasize the importance of continued recording and model retraining when providing personalized models for individuals with progressive speech impairments.  
Amassing speech recordings during the early stage of the disease may be unnecessary if it is solely to improve future recognition when speech becomes more severe.
Our finding that A-SI models can perform similar or even better than  un-updated personalized models suggests that A-SI models may be a worthwhile option if re-recording is not possible.

These experiments were performed on a relatively small cohort of speakers. In the future, we plan to extend to more speakers and include other etiologies that lead to degenerating speech. We are currently recording a more controlled, longitudinal dataset with additional participants.


\bibliography{main}

\end{document}